\documentclass[aps,prl,showpacs,twocolumn]{revtex4}
\usepackage{bbm}
\usepackage{mathrsfs}
\usepackage{epsfig,psfrag}
\usepackage{amsmath,amsfonts,amssymb}

\newcommand{\Ecal}{\mathcal E}
\newcommand{\Mcal}{\mathcal M}

\newcommand{\xbf}{\mathbf {x}}

\newcommand{\pbf}{\mathbf {p}}
\newcommand{\qbf}{\mathbf {q}}

\newcommand{\sbf}{\mathbf {s}}
\newcommand{\ubf}{\mathbf {u}}
\newcommand{\nbf}{\mathbf {n}}

\newcommand{\pd}{\partial}
\newcommand{\til}[1]{\tilde{#1}}

\begin{document}

\preprint{draft}

\title{Geometry and Spectrum of Casimir Forces}

\author{Rauno B\"uscher and Thorsten Emig}

\affiliation{Institut f\"ur Theoretische Physik, Universit\"at zu
K\"oln, Z\"ulpicher Stra\ss e 77, D-50937 K\"oln, Germany}

\date{\today}

\begin{abstract}
  We present a new approach to the Helmholtz spectrum for arbitrarily
  shaped boundaries and general boundary conditions. We derive the
  boundary induced change of the density of states in terms of the
  free Green's function from which we obtain non-perturbative
  results for the Casimir interaction between rigid surfaces. As an
  example, we compute the lateral electrodynamic force between two
  corrugated surfaces over a wide parameter range. Universal behavior,
  fixed only by the largest wavelength component of the surface shape,
  is identified at large surface separations, complementing known
  short distance expansions which we also reproduce with high
  precision.
\end{abstract}


\maketitle

The famous question ``Can one hear the shape of a drum?'' was posed in
1966 by Kac, illustrating the problem to deduce the shape of a region
from the knowledge of its resonance spectrum \cite{Kac66}. It was
answered negatively \cite{Gordon+92} but the difficulty to
characterize the distribution of eigenfrequencies of the Helmholtz
wave equation in arbitrary geometries remains. This is particularly
highlighted by the long-lasting efforts for chaotic (quantum)
billiards which are described in two dimensions also by the wave
equation \cite{ChaosBook}. The same problems occur for Casimir
interactions in three dimensions \cite{Casimir48} which are of much recent interest
\cite{Bordag+01}, mainly due to the advent of improved experimental
techniques yielding the force between two parallel plates
\cite{Bressi+02} and between a plate and a sphere \cite{Lamoreaux00}.
Naturally, the question arises to what extent the Casimir force
characterizes the shape of the interacting objects. A large body of
theoretical work, including proximity and pairwise additivity
approximations \cite{Bordag+01}, semi-classical approaches based on
Gutzwiller's formula \cite{Gutzwiller+71,Schaden+98}, a multiple
scattering expansion \cite{Balian-Duplantier+78-1}, perturbative
methods \cite{Emig+01+03} and recently proposed approximations
from classical ray optics \cite{Jaffe+04}, has been used to compute
Casimir forces. However, these approaches either neglect diffraction
or are limited to small smooth deformations, and reliable general expressions
are not known even for simple geometries.  In this Letter, we present
a formula for the density of states which is formulated in terms of
the free Green's function at the boundaries only. We demonstrate its
applicability by computing the lateral Casimir force between two
corrugated surfaces which has been studied in a recent experiment
\cite{Chen+02a}. We find that the surface's shape can be deduced from
the force at short distances whereas at large separations universality
prevails.

The electrodynamic Casimir energy of two disconnected metallic
surfaces $S_\alpha$, $\alpha=1,2$, is determined by the {\it change}
in the photon density of states (DoS) $\delta \rho (k)=
\rho(k)-\rho_0(k)$ caused by moving the surfaces from infinity to a
finite distance in vacuum where $\rho_0(k)$ is the DoS for infinitely
separated surfaces. Thus $\delta \rho(k)$ contains neither volume
terms nor single surface contributions but measures only {\it changes}
in geometry by moving the surfaces rigidly.  We consider the
Helmholtz equation
\begin{equation}
(\nabla^2 + k^2) \phi(\xbf) = 0 
\end{equation}
for a scalar field $\phi$ in the three connected regions separated by
surfaces $S_\alpha$ on which $\phi$ fulfills Dirichlet (for
transversal magnetic modes, TM) or Neumann (for transversal electric
modes, TE) boundary conditions.  The total DoS $\rho(k)$ is then given
by the sum of the DoS for the three isolated regions.  For simplicity,
we have assumed that the surface geometry allows for a separation of
the electromagnetic field into TE and TM modes which is possible for a
large class of geometries as, e.g., for uniaxially deformed surfaces
\cite{Emig+01+03}. For both types of modes the Casimir energy is obtained
as an integral over $\delta \rho(k)$. Since the DoS is more regular
along the imaginary axis, it is useful to shift the integration to
that axis and to go over from a Minkowskian to a Euclidean formulation
by a Wick rotation. Then the Casimir energy becomes
\begin{equation}\label{eq:general-energy}
\Ecal\:=\:\frac{\hbar c}{2}\int_0^\infty dq_0\,q_0\,\delta\tilde\rho(q_0) \,
\end{equation}
with $\delta \tilde\rho(q_0)\equiv -\delta\rho(iq_0)$. Our main general
result is the trace formula
\begin{equation}\label{eq:trace-formula}
\delta\tilde\rho(q_0)\:=\:-\frac{1}{\pi}\frac{\pd}{\pd q_0}\,
\text{Tr}\ln\left(\Mcal\Mcal_\infty^{-1}\right) \, ,
\end{equation}
where the matrix operator $\Mcal$ is given by the Euclidean free
Green's function
$G_0(\xbf,\xbf';q_0)=e^{-q_0|\xbf-\xbf'|}/(4\pi|\xbf-\xbf'|)$
evaluated at the surfaces $S_\alpha$ only. Explicitly, if the surfaces
$S_\alpha$ are represented by 3D vectors $\sbf_\alpha(\ubf)$ with 2D
local coordinates $\ubf$, then
$\Mcal_{\alpha\beta}(\ubf,\ubf';q_0)=G_0(\sbf_\alpha(\ubf),
\sbf_\beta(\ubf');q_0)$ for Dirichlet conditions and
$\Mcal_{\alpha\beta}(\ubf,\ubf';q_0)=\pd_{\nbf_\alpha(\ubf)}
\pd_{\nbf_\beta(\ubf')} G_0(\sbf_\alpha(\ubf),\sbf_\beta(\ubf');q_0)$
for Neumann conditions with $\pd_{\nbf_\alpha}$ the surface normal
derivative pointing into the region between the surfaces.  The trace
in Eq.~(\ref{eq:trace-formula}) is performed over the 2D coordinates
and the discrete surface indices.  $\Mcal_\infty^{-1}$ is the
functional inverse of $\Mcal$ for infinite surface separation.  A
formally similar formula for the Casimir DoS has been derived by
Balian and Duplantier in terms of a different matrix operator which
describes surface scatterings \cite{Balian-Duplantier+78-1}. In the
spectral theory of quantum scattering an expression of the form of
Eq.(\ref{eq:trace-formula}) is known as Krein-Friedel-Lloyd formula
which, however, applies to the S-matrix of potential scatterers
\cite{ChaosBook}.  An important advantage of
Eq.(\ref{eq:trace-formula}) is that it yields directly the regularized
variation of the DoS which is free of distance independent divergences
which would seriously hamper any numerical evaluations.  To set an
example for the applicability of our trace formula to the Casimir
effect in non-trivial geometries, we compute the {\it lateral} Casimir
force between corrugated surfaces which is especially sensitive to
geometry.  For that purpose, we employ a previously developed
numerical algorithm \cite{Emig03,buescher+emig+04-1} which provides a
fast convergent result for the trace. It is important to note that the
lateral force does not arise from a change of the mean surface
separation (yielding the normal force) but from a lateral shift of the
boundaries, and thus requires a careful regularization of the energy.
   
At first, we give a brief survey of the steps which lead to
Eq.~(\ref{eq:trace-formula}).  We consider the Gaussian action
$S=\frac{1}{2}\int d^4 X (\nabla \phi)^2$ in 4D Euclidean space-time
to quantize the modes of the electromagnetic field in the regions
which are separated by the surfaces $S_\alpha$ on which Dirichlet
or Neumann boundary conditions hold. Path integral techniques with
delta functions enforcing the boundary conditions have been used to
compute the (free) energy of constrained systems
\cite{likar,Emig+01+03}. The same approach, however, can be also used
to study correlations \cite{Hanke+02}. The modified correlations
$G(\xbf,\xbf';q_0)=\langle \phi_{q_0}(\xbf)\phi_{-q_0}(\xbf')\rangle$
in the presence of boundaries can be computed exactly in the present
case of a quadratic action. If $\xbf$ and $\xbf'$ denote two positions
which are located both in the {\it same} region, one finds
\begin{eqnarray}
\label{eq:boundary-green}
G(\xbf,\xbf';\!q_0)\!\!&-&\!\!G_0(\xbf,\xbf';\!q_0)
\!=\! -\!\!\sum_{\alpha\beta} 
\!\int \!\! d\ubf \, d\ubf' \,
G_0(\xbf,\sbf_\alpha(\ubf);\!q_0) \nonumber \\
&\times&\Mcal^{-1}_{\alpha\beta}(\ubf,\ubf';\!q_0)\, 
G_0(\xbf',\sbf_\beta(\ubf');\!q_0) \, ,
\end{eqnarray}
where $G_0$ is Green's function in unbounded space.  $\Mcal^{-1}$ is
the functional inverse, taken with respect to $\ubf$, $\ubf'$ and
$\alpha$, $\beta$, of the operator $\Mcal$ defined after
Eq.~(\ref{eq:trace-formula}), see above.  For a fixed region, the DoS
on the imaginary axis is related to the Euclidean Green's function by
$\rho(iq_0) = (2q_0/\pi) \int d^3 \xbf\, G(\xbf,\xbf;q_0)$ where the
integration extends over the given region. We are actually interested
in the sum of the DoS's of the three regions into which free space is
divided by the surfaces with the bulk contribution subtracted. Since
Eq.~(\ref{eq:boundary-green}) holds for every region, we obtain the
change $\delta \rho(iq_0)$ by integrating the r.h.s. of
Eq.~(\ref{eq:boundary-green}) with $\xbf'=\xbf$ over unbounded space.
Explicit integration is enabled by the simple form of
Eq.~(\ref{eq:boundary-green}) with $\xbf$ occurring only in the free
Green's function.  Performing the integration both with $\Mcal$ and
$\Mcal_\infty$ and taking the difference of the two results, we obtain
Eq.~(\ref{eq:trace-formula}). 

\begin{figure}[b]
\vspace*{-.3cm}
\includegraphics[width=.85\linewidth]{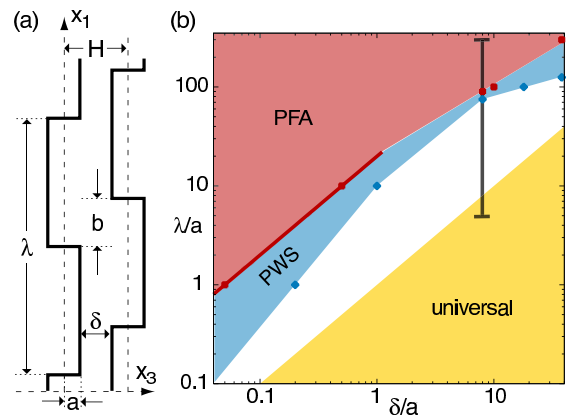}
\vspace*{-.3cm}
\caption{\label{fig1} 
  (a) Geometry consisting of two parallel plates with laterally
  shifted uniaxial rectangular gratings. (b) Approximate validity
  ranges of proximity force (PFA) and pairwise (PWS) approximations
  and sector of asymptotic universality for the lateral Casimir
  force as estimated from Figs.\ref{plot1} and \ref{plot2}.}
\end{figure}

\begin{figure}[ht]
\includegraphics[width=.85\linewidth]{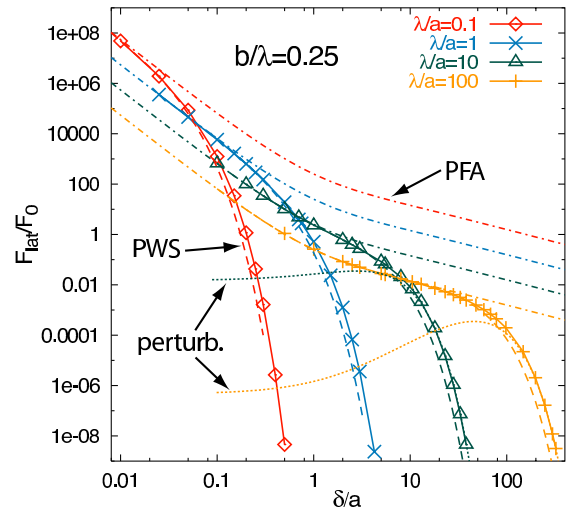}
\vspace*{-.3cm}
\caption{\label{plot1} 
  Lateral force $F_{\rm lat}$ at $b=\lambda/4$
  obtained from Eq.~(\ref{eq:force_lat}) for the geometry of
  Fig.~\ref{fig1}(a) as function of the surface gap $\delta$ (solid
  curves).  $F_{\rm lat}$ is measured in units of the normal force
  $F_0$ between flat surfaces with $a=0$.  Plotted are also the
  proximity force (PFA, dash-dotted curves) and pairwise summation
  (PWS, dashed curves) approximations, and the perturbative result
  $F_{\rm pt}$ for sinusoidal profiles (dotted curves)
  \cite{Emig+01+03}.}
\end{figure}

\begin{figure}[tb]
\includegraphics[width=.85\linewidth]{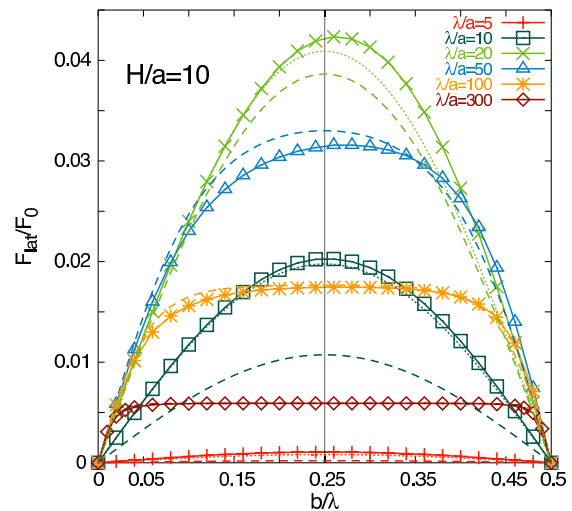}
\vspace*{-.3cm}
\caption{\label{plot3} 
  Shape dependence of $F_{\rm lat}$ on the lateral surface shift $b$
  at fixed distance $H=10a$ for different corrugation lengths. The
  dashed and the dotted curves represent the PWS and the full
  perturbative result for sinusoidal profiles with arbitrary
  $H/\lambda$ \cite{Emig+01+03}, respectively.}  \vspace*{-.5cm}
\end{figure}

In the following, we consider the geometry depicted in
Fig.~\ref{fig1}(a). It consists of two surfaces with uniaxial
rectangular gratings along the $x_1$ axis with equal amplitude $a$ and
wavelength $\lambda$. The surfaces are laterally shifted by $b$ and
have a mean separation $H$, yielding a minimal gap $\delta=H-2a$.  The
Casimir energy of these surfaces at fixed $H$ has to be a periodic
function of $b$, and it should be minimal if the surface area with
minimal surface distance is maximal, i.e., for $b=\lambda$. This leads
to a lateral force $F_{\rm lat} = -\partial \Ecal/\partial b$.
Performing the frequency integration by parts in
Eq.~(\ref{eq:general-energy}) with Eq.~(\ref{eq:trace-formula}), we
find
\begin{equation}
\label{eq:force_lat_with_M}
F_{\rm lat} = -\frac{\hbar c}{2\pi} \int_0^\infty \!\!\! dq_0 \,
{\rm Tr} \left(
\Mcal^{-1} \partial_b \Mcal - \Mcal_\infty^{-1} \partial_b \Mcal_\infty
\right) \, .
\end{equation}
For periodic geometries as the one considered here, the trace can be
computed with the technique introduced in
\cite{Emig03,buescher+emig+04-1}. We Fourier transform
$\Mcal_{\alpha\beta}(\ubf,\ubf';q_0)$ with respect to $\ubf$ and
$\ubf'$ so that use can made of the periodicity along $x_1$ and
translational invariance along $x_2$, allowing for the representation
\begin{equation}\label{eq:Bloch}
\begin{split}
  &\til{\Mcal}_{\alpha\beta}\left(\pbf,\qbf;q_0\right)\:=\:
  (2\pi)^2\delta\left(p_2 + q_2 \right)\\
  &\times\sum_{m=-\infty}^\infty \delta\left(p_1+q_1+2\pi
    m/\lambda\right)\, N_m^{\alpha\beta}\left(q_1,q_2;q_0\right) \, ,
\end{split}
\end{equation} 
which defines the $2\times2$ matrices $N_m$ that can be computed
analytically and are given in \cite{buescher+emig+04-1} for the
geometry of Fig.\ref{fig1}(a). The trace in
Eq.(\ref{eq:force_lat_with_M}) is more efficiently obtained if it is
restricted to momenta $q_1$ in the interval $[0,2\pi/\lambda)$. This
is possible after a rearrangement of the elements of $\til{\Mcal}$ so
that it has block-diagonal form where the blocks are numbered by the
continuous index $q_1$ in $[0,2\pi/\lambda)$ and, for fixed $q_0$,
$q_2$, have matrix elements
$B_{kl}^{\alpha\beta}(q_1,q_2;q_0)=N_{k-l}^{\alpha\beta}(q_1+2\pi
l/\lambda,q_2;q_0)$ for integer indices $k$,
$l=-\infty,\ldots,\infty$. Physically, a block with index $q_1$
couples only waves whose momenta differ from the Bloch momentum $q_1$
by integer multiples of $2\pi/\lambda$. Thus, in analogy to Bloch's
theorem, the original problem has separated into decoupled subproblems
at a fixed $q_1$ in the interval $[0,2\pi/\lambda)$, and the total
trace in Eq.(\ref{eq:force_lat_with_M}) is given by the sum over the
traces of the subproblems. This fact can be expressed by defining the
function $g(q_1,q_2;q_0)=\text{tr}(B^{-1}\pd_bB
-B^{-1}_\infty\pd_bB_\infty)$ where the trace runs over the indices
$k$, $l$, $\alpha$, $\beta$ of $B$, and $B_\infty$ is the analog of
$\Mcal_\infty$, i.e., $B$ for $H\to\infty$.  The lateral Casimir force
is then given by
\begin{equation}
\label{eq:force_lat}
F_{\rm lat} = - \frac{\hbar c}{8\pi^2}
\int_0^\infty \! dq\, q \int_0^{2\pi/\lambda} \! dq_1 \, g(q_1,q_2;q_0) \, ,   
\end{equation}
with $q=\sqrt{q_0^2+q_2^2}$. Since the matrices $N_m$ are known
analytically \cite{buescher+emig+04-1}, the same applies to $B$,
$B_\infty$, and the derivative with respect to $b$ can be easily
computed. However, for a non-perturbative treatment, the inversions of
$B$, $B_\infty$ have to be implemented numerically.  This is enabled
by a truncation of the matrix $B$ at a fixed order $M$ so that the
function $g$ is replaced by $g_M$ which is defined as $g$ above but
with the trace running over $k,l=-M\dots M$ only. $F_{\rm lat}$
follows then from a numerical integration over $g_M$ in
Eq.~(\ref{eq:force_lat}) for a sequence of fixed $M$ and a subsequent
extrapolation to $M\to \infty$. For the results shown below, we have
chosen $M$ between $13$ and $37$ with the larger $M$ used at smaller
separations $\delta$. This is physically consistent with the fact that
with increasing separation smaller integer multiples of $2\pi/\lambda$
around the Bloch momenta have to be considered.  It should be stressed
that the above analysis is independent of the boundary conditions
which, however, change $\Mcal$.  Thus the electromagnetic Casimir
force is the sum of $F_{\rm lat}$ for Dirichlet and Neumann boundary
conditions, respectively, leading to the results for $F_{\rm lat}$
summarized in Figs.~\ref{plot1}-\ref{plot3}.


\begin{figure}[ht]
\includegraphics[width=.85\linewidth]{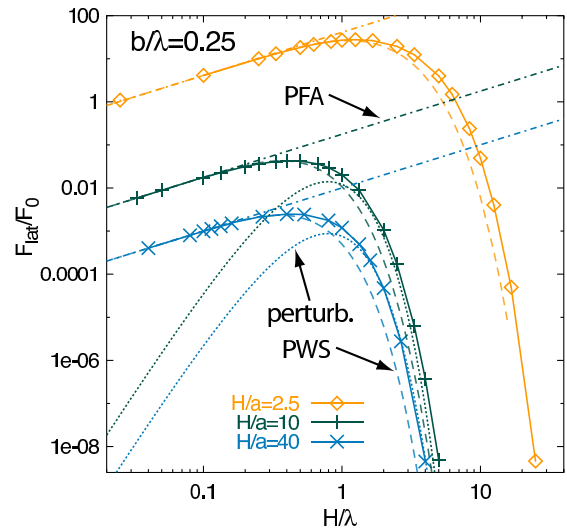}
\vspace*{-.3cm}
\caption{\label{plot2} 
  Dependence of $F_{\rm lat}$ (solid curves) on the corrugation length
  $\lambda$ at fixed mean distances $H$. Shown are also the
  approximations (PFA, PWS) for short distances and the perturbative
  result $F_{\rm pt}$ for $a \ll H$, $\lambda$.}
\vspace*{-.5cm}
\end{figure}

At short surface distances approximative methods can be employed and
it is instructive to compare their predictions to our findings.  To
begin with, the proximity force approximation \cite{Bordag+01} yields
the lateral force $F_{\rm
  PFA}=[2\Ecal_0(H)-\Ecal_0(H-2a)-\Ecal_0(H+2a)]/\lambda$ for
$0<b<\lambda/2$ with $\Ecal_0(H)=-(\pi^2/720)\hbar c/H^3$ since it
sums the flat plate energy $\Ecal_0$ at the local distance normal to
the surfaces. $F_{\rm PFA}$ changes sign at $b=\lambda/2$
discontinuously. A different approximation consists in the pairwise
summation (PWS) of Casimir-Polder potentials.  Although strictly
justified for rarefied media only, it is often also applied to metals,
using the two-body potential $U(r)=-(\pi/24)\hbar c/r^7$ whose
amplitude is chosen as to reproduce the correct result for flat ideal
metal plates \cite{Bordag+01}. It yields the lateral force $F_{\rm
  PWS}=-\frac{\partial}{\partial b} \int_{V_l} d^3 \xbf \int_{V_r} d^3
\xbf' U(|\xbf-\xbf'|)$ with $V_l$ and $V_r$ denoting the semi-infinite
regions to the left and right of the two surfaces in
Fig.\ref{fig1}(a), respectively. To compute $F_{\rm PWS}$, we have
first differentiated analytically with respect to $b$ and then
performed the remaining integrals numerically. Fig.\ref{plot1} shows
our results for the amplitude of the lateral force, measured relative
to the normal force $F_0$ between flat plates at the same $H$, over
more than 4 orders of magnitude for the gap $\delta$ together with the
two approximations.  For small $\delta$, both approximations agree and
match our results. Beyond $\delta \gtrsim \lambda/20$ the PFA starts
to fail since it does not capture the exponential decay of $F_{\rm
  lat}$ for increasing $\delta$. The PWS approach has a slightly
larger validity range [cf.~Fig.\ref{fig1}(b)] and reproduces the
exponential decay. However it deviates by at least {\it one order of
  magnitude} from $F_{\rm lat}$ for $\delta \gtrsim 2.5 \lambda$, see
Fig.\ref{plot1}. Thus it is important that in the asymptotic limit of
large surface gaps, one can expect a universal behavior of the force
which is independent of the detailed shape of the surface corrugation.
Precisely this expectation is strikingly confirmed when we compare our
results to the perturbative expression for the lateral force for {\it
  sinusoidally} shaped surfaces (with amplitude $a_0$ and wavelength
$\lambda$) \cite{Emig+01+03},
\begin{equation}
\label{eq:f-pert}
F_{\rm pt}=\frac{8\pi^3\, \hbar c}{15} \frac{a_0^2}{\lambda^5 H}\,
\sin\left(\frac{2\pi}{\lambda} \, b\right) e^{-2\pi H/\lambda} \, ,
\end{equation}
which is expected to hold for $a_0\ll \lambda \ll H$. By considering
the lowest harmonic of the rectangular corrugation, which corresponds
in Eq.(\ref{eq:f-pert}) to $a_0=4a/\pi$, we find excellent agreement
between $F_{\rm pt}$ and our results for the geometry of
Fig.\ref{fig1}(a) for large distances $\delta \gtrsim \lambda$, see
Fig.\ref{plot1}. Higher harmonics describing short scale surface
structure are irrelevant for the asymptotic Casimir interaction. This
is particularly highlighted by the dependence of the force on the
lateral shift $b$ shown in Fig.\ref{plot3} for $H=10a$ and $\lambda$
varying in the interval indicated by the bar in Fig.\ref{fig1}(b).
With changing $\lambda$, three regimes can be identified. For $\lambda
\gg H$, the force profile resembles almost the rectangular shape of
the surfaces, and the PWS approach yields consistent results. For
decreasing $\lambda$, yet larger than $H$, the force profile becomes
asymmetric with respect to $b=\lambda/4$ and more peaked, signaling
the crossover to the universal regime for $\lambda \lesssim H$ where
the force profile is sinusoidal. In the latter case, for not too small
$\lambda/a \approx 10$, our results for $F_{\rm lat}$ indeed agree
well with the perturbative result for arbitrary $H/\lambda$ of
Ref.\cite{Emig+01+03}, cf.~Fig.\ref{plot3}.  We note that the PWS
approach fails to predict the asymmetry of the force profile, and the
PFA even predicts no variation at all with $b$ in the range of
Fig.\ref{plot3}. Finally, we observe a non-monotonous change of the
lateral force with $H/\lambda$ when $H/a$ is kept fixed, see
Fig.\ref{plot2}. After a linear increase of $F_{\rm lat}/F_0$,
described by the PFA, the force shows a maximum beyond which it decays
exponentially. For small amplitudes, $a/H=1/40$ in Fig.\ref{plot2},
the position of the maximum at $H\approx 0.4\lambda$ is again in good
agreement with the full perturbative result \cite{Emig+01+03}. With
increasing $a/H$, the maximum is shifted towards smaller wavelengths.

In conclusion, we have derived a formula for the change of the
Helmholtz spectrum by arbitrarily shaped boundaries. From
non-perturbative results based on this formula, we argue that the
lateral Casimir force between {\it any} two uniaxially corrugated
surfaces of equal wavelength and amplitude is described by
Eq.(\ref{eq:f-pert}) for $H$ much larger than the wave length. We note
that the validity range for this universal behavior is not fully realized
in the experiment of \cite{Chen+02a} with $H/a\approx 10$ and
$\lambda/a\approx 55$ when $a$ is set to the geometric mean of the
distinct amplitudes of the experiment. It would be interesting to
probe this universality in experiments with different shapes
at larger ratios $H/\lambda$. We studied surface deformations with a
bounded spectrum. Stochastic surface roughness does not have this
feature, and we then expect a different asymptotic behavior. We note
that our approach yields also the non-integrated DoS and can be easily
used in any space dimension, and also for closed boundaries. This
might be of importance for applications to chaotic systems as quantum
billiards. At finite temperatures, the DoS has to be integrated with a
saw teeth-like weight \cite{Balian-Duplantier+78-1}.  Finally,
material properties of the surfaces can be included in our approach in
form of non-local boundary conditions \cite{buescher+emig+04-2}.

This work was supported by the Deutsche Forschungsgemeinschaft through
the Emmy Noether grant No. EM70/2-3.

\end{document}